# Unraveling Lithium Dynamics in Solid Electrolyte Interphase: From Graph Contrastive Learning to Transport Pathways

Qiye Guan[*] and Yongqing Cai[*]

*Institute of Applied Physics and Materials Engineering, University of Macau, Taipa, Macau, China*

[*]*Corresponding authors:* qiye.guan@connect.um.edu.mo; yongqingcai@um.edu.mo

**Abstract**

Fast lithium transport across the solid-state electrolyte (SSE) /lithium metal anode interface is critical for high-performance all-solid-state batteries. Uncovering the complex lithium dynamics governed by diverse local environments in the solid electrolyte interphase (SEI) is fundamental for performance optimization. However, a general framework for characterizing these distinct local environments and the associated transport mechanisms remains lacking. Here, we develop GET-SEI, a general framework that discovers local atomic environments without predefined labels through Graph contrastive learning (GCL), models lithium transition kinetics via Extended dynamic mode decomposition (EDMD), and quantifies reactive lithium flux through Transition path theory (TPT). Applied to different SSE/Li systems, including sulfides ($Li_6PS_5Cl$/Li, $Li_{10}GeP_2S_{12}$/Li) and oxides ($Li_7La_3Zr_2O_{12}$/Li), the GET-SEI reveals dominant transport pathways and kinetic bottlenecks in each system, providing quantitative metrics for evaluating lithium transport efficiency. As novel high-performance SSEs continue to emerge, GET-SEI offers a widely applicable, interpretable tool for targeted SEI engineering.

**Introduction**

The development of all-solid-state batteries has reached a critical stage, driven by the promise of high energy density and improved safety.[1] Substantial efforts have focused on discovering solid-state electrolytes (SSEs) with high ionic conductivity, compatibility with lithium metal, and electrochemical stability.[2-4] At the same time, understanding the solid electrolyte interphase (SEI) is equally important, because this interphase governs interfacial lithium transport, dendrite formation, and parasitic side reactions.[5-14] Given the abundant and diverse local environments in amorphous SEIs,[15-17] characterizing lithium dynamics across these environments is essential for designing interphases that enable rapid and stable interfacial transport.[8,18]

Experimentally, techniques such as X-ray photoelectron spectroscopy (XPS), scanning electron microscopy (SEM), and operando X-ray computed tomography have provided direct evidence of SEI formation and dendrite evolution.[10,13,15] However, lithium dynamics at the microscopic level within the SEI remain insufficiently resolved. Molecular dynamics and Monte Carlo simulations can capture aspects of SEI formation and structural evolution[19-22], but extracting interpretable lithium transport dynamics from these high-dimensional trajectories remains challenging. Analyses based on representative intermediates offer useful qualitative insights, yet often oversimplify dynamic transport processes.[5,19,23,24] Therefore, a computationally efficient and generalizable framework that captures complex lithium dynamics is needed for predictive SEI modeling and rational interphase design.

Although deep kinetic learning methods[25,26] (e.g., VAMPnets[27,28] and graph dynamical networks[29]) can capture complex transition patterns, they typically provide latent coordinates with limited physical interpretability.[30-32] In SEI systems, where lithium transport is controlled by diverse and system-specific local environments, this limitation prevents linking learned dynamics to physically meaningful local environments. If local environments are identified as interpretable states, their transitions can be projected onto reduced Koopman models[33,34] via methods like extended dynamic mode decomposition (EDMD)[35] to obtain tractable, approximately linear dynamics for quantitative analysis. Identifying representative local chemical environments and quantifying their similarity is therefore the central challenge. Graph-based deep learning provides a natural representation of atomistic structures and has been successfully applied to universal force-field development[36,37] and structure-property prediction[38-41]. For state discovery and classification, graph contrastive learning (GCL) offers an efficient self-supervised framework for learning similarity across multiple graph scales.[42,43] This makes GCL particularly suitable for abundant Li-centered micro-environment graphs in SEI systems, enabling robust and accurate discovery of physically meaningful local-state relationships.

Here, we establish GET-SEI, a general pipeline that integrates graph contrastive learning (GCL), extended dynamic mode decomposition (EDMD), and transition path theory (TPT)[44,45] to model the complex dynamics of lithium atoms in the SEI. Applied to representative SEI systems, including $Li_6PS_5Cl$/Li (LPSCl/Li), $Li_{10}GeP_2S_{12}$/Li (LGPS/Li) and $Li_7La_3Zr_2O_{12}$/Li (LLZO/Li), GET-SEI demonstrates both computational efficiency and cross-system generalizability. Unlike black-box latent models, GET-SEI yields mechanistically interpretable, state-based representations that are directly linked to local structural features. EDMD and subsequent TPT analysis further identify rate-limiting transitions and dominant escape pathways, providing quantitative insight into lithium conduction and a practical framework for SEI design and evaluation.

## Results

### Identify local environments in SEI through graph contrastive learning

Unlike periodically ordered inorganic SSEs or lithium metal phases, amorphous SEIs and their associated lithium transport dynamics are intrinsically difficult to model because of the abundance and diversity of local environments. Each mobile lithium atom can exhibit a distinct local fingerprint that is not fully described by bond topology or static atomic coordinates alone. To address this complexity, we represent each Li-centered neighborhood as a local graph, $\mathcal{G} = (V, E)$, where vertices $V$ denote atoms and edges $E$ encode local connectivity. By incorporating atomic node features, such as coordination descriptors, average neighbor distances, and local positional statistics, these graphs capture subtle but physically relevant environmental variations.

We then employ graph contrastive learning (GCL) to compare and classify these local environments. As shown in **Fig. 1a**, input graphs are constructed from neural-network molecular dynamics (NNMD) trajectories. (**see Methods for details**) The core principle of contrastive learning is to maximize agreement between representations of the same sample under suitable augmentations.[46] Specifically, each graph $\mathcal{G}$ is transformed into two correlated views, $\mathcal{G}^{(1)}$ and $\mathcal{G}^{(2)}$, using random edge dropout and random node-feature masking. These augmented graphs are encoded by a scalar-invariant graph attention network (GAT)[42,47] with two multi-head attention layers. Each node is represented by scalar attributes and message passing aggregates neighboring information in a non-directional manner. The encoder output is projected into a 32-dimensional embedding space.

To quantify graph similarity, we train the GCL model using the information Noise Contrastive Estimation loss ($\mathcal{L}_{infoNCE}$)[42]:

$$\mathcal{L}_{infoNCE} = -\log \frac{\exp(f(z^{(1)}, z^{(2)})/\sigma)}{\sum_{j=1}^{N} \exp(f(z^{(1)}, z^{(2)})/\sigma)} \quad (1)$$

Where $z^{(1)}$ and $z^{(2)}$ denote the embeddings of the same Li-centered local graph from the two augmented views, $\mathcal{G}^{(1)}$ and $\mathcal{G}^{(2)}$, respectively. $f$ is the cosine similarity function, $N$ is batch size, and $\sigma$ is the parameter controlling the sharpness of the contrastive distribution. This self-supervised objective organizes the embedding space such that similar local environments are mapped nearby, while dissimilar environments are separated. The learned embeddings are then clustered using a Gaussian mixture model (GMM) with an optimally selected number of components, enabling identification of dominant local-environment states in SEI systems.

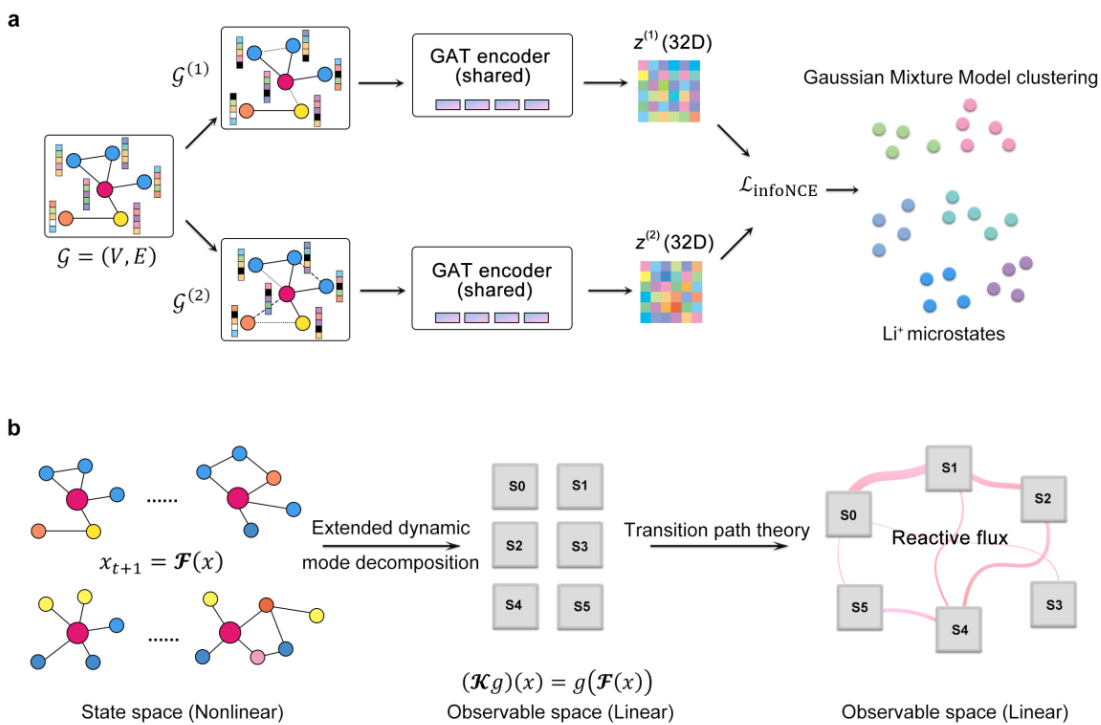

**Fig. 1 GET-SEI framework for analyzing lithium dynamics in SEIs. a**, Graph contrastive learning model in finding local environments in SEIs. **b,** Modeling of lithium dynamics through extended dynamic mode decomposition and lithium flux analysis through transition path theory. The S0 to S5 denotes lithium states identified by graph contrastive learning.

**Modeling Lithium dynamics through EDMD and TPT**

Having classified local environments in the SEI via GCL, we next apply the Koopman theory[33] to represent the generally nonlinear dynamics of SEI systems in a linear framework. Instead of tracking how states evolve, this approach monitors how observables (scalar functions of states) evolve. For a dynamic system with flow map $\mathcal{F}$, the state update follows:

$$x_{t+1} = \mathcal{F}(x_t) \quad (2)$$

Where $x \in \mathcal{M}$ lies in the original high-dimensional chemical space, and full atomic velocities/positions need to be considered. Directly modeling dynamics in $\mathcal{M}$ is computationally challenging, so we introduce Koopman operator $\mathcal{K}$, which acts linearly on observables $g: \mathcal{M} \to \mathbb{R}$. Specifically, $\mathcal{K}$ maps the observable $g$ to a new function $\mathcal{K}g$ such that evaluating the new function at the current state $x$ gives the observable's value at the next state:

$$(\mathcal{K}g)(x) = g(\mathcal{F}(x)) \quad (3)$$

In this way, a finite-dimensional nonlinear dynamical system is equivalently described by a linear, infinite-dimensional operator $\mathcal{K}$ acting on observables.

Using EDMD[35], we estimate a finite-dimensional approximation of $\mathcal{K}$ from trajectory data. (**details in Methods**) The equilibrium population of each state is obtained from the left eigenvector of $\mathcal{K}$ with eigenvalue $\lambda = 1$. The Koopman matrix further provides transition probabilities among Li local-environment states. We then compute the continuous-time rate matrix $R$ via $\mathcal{K} = \exp(R \cdot \tau)$:

$$R = \ln\mathcal{K}/\tau \quad (4)$$

Where $\tau$ is the lag time. To move beyond rate magnitudes and resolve mechanistic pathways, we perform analysis through TPT[44]. The reactive flux $f_{ij}^+$ from states $i$ to $j$ is:

$$f_{ij}^+ = \pi_i \cdot q_i^- \cdot \mathcal{K}_{ij} \cdot q_j^+ \quad (5)$$

Where $\pi_i$ is the equilibrium population at state $i$, and $q_i^-$, $q_j^+$ are backward/forward committor probabilities, respectively. This flux quantifies productive lithium transport between distinct local environments in the SEI and enables identification of dominant pathways and kinetic bottlenecks.

**Case study in LPSCl/Li system**

We first applied GET-SEI to characterize lithium dynamics in the LPSCl/Li interphase. Experimentally verified SEI thicknesses span hundreds of nanometers.[15,20] To model nanoscale SEI formation, we performed NNMD simulations (**Supplementary Table 1**). The LPSCl/ Li interface model was built from (100) surfaces with a total of 2960 atoms. A 1.0 ns NNMD simulation was then carried out in the NVT ensemble at 300 K to sample interphase evolution. (**Fig. 2a**) Li-centered local-environment graphs were subsequently constructed from the NNMD trajectories. (**Methods section**)

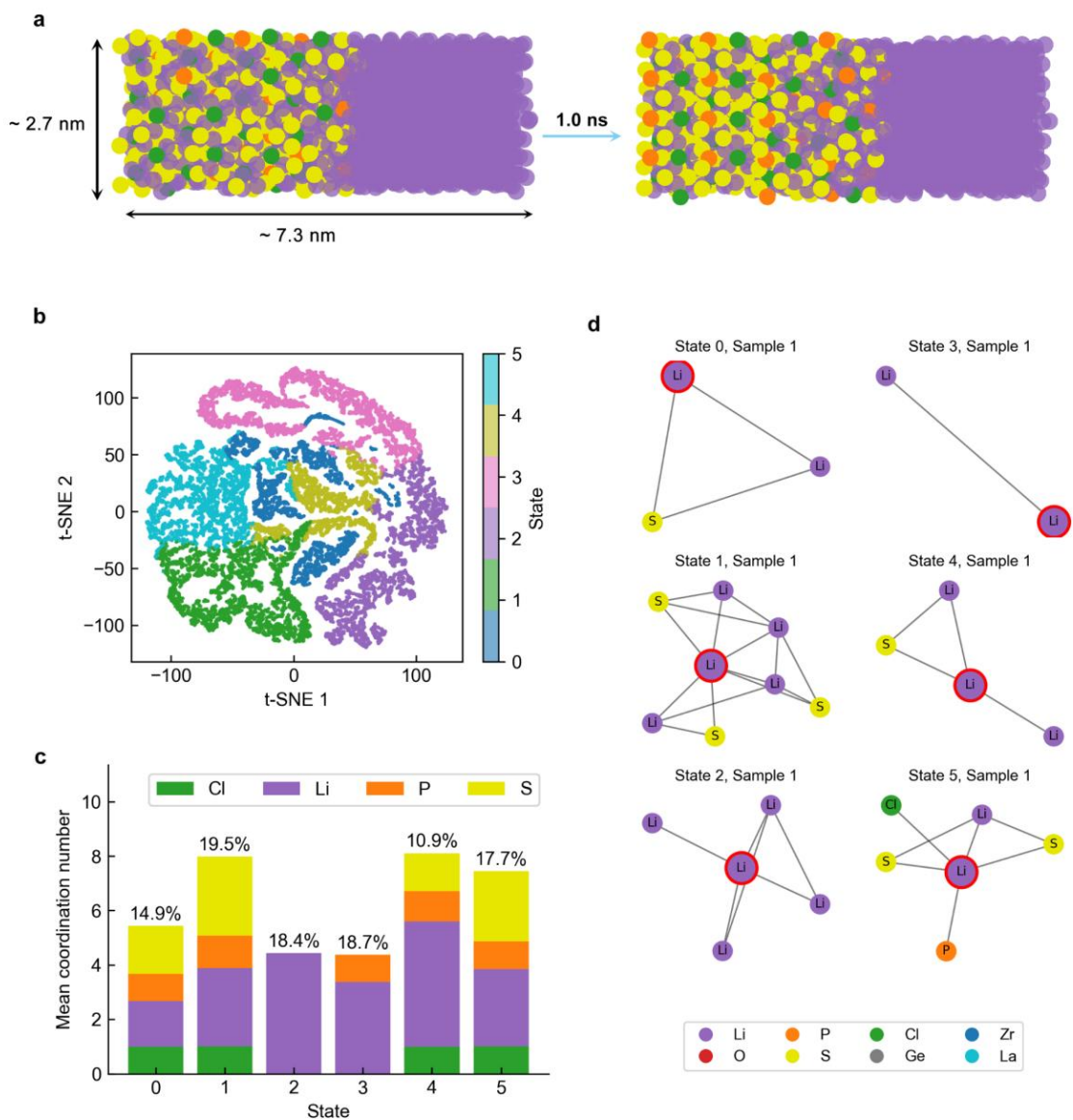

**Fig. 2 Representative local environments identified through graph contrastive learning in LPSCl/Li. a,** LPSCl/Li model used for molecular dynamics simulation before and after 1.0 ns. **b**, t-SNE visualization of the embedded space in 2 dimensions. **c**, Mean coordination number of six states (S0-S5) in LPSCl/Li under a 4.0 Å

cutoff. **d**, Representative graphs of six states (S0-S5) in LPSCl/Li system. The center lithium is denoted with a red circle.

To obtain robust local-environment classification while avoiding both over-fragmented and overly coarse observables, we selected six states (S0-S5), which yields a Silhouette score of 0.29. (**Fig. 2b, Supplementary Table 2, Supplementary Note 1**) These states exhibit distinct coordination environments and represent dominant structural motifs during interphase evolution. (**Fig. 2c-2d**) Additional representative samples in the embedding space are provided in **Supplementary Fig. 1**. We then linearized the lithium dynamics in the original chemical space using the Koopman operator $\mathcal{K}$ with a lag time of 80.0 ps, identifying five dominant dynamical processes in the LPSCl/Li SEI system. (**Fig. 3a**)

The escape-rate matrix (**Fig. 3b**) quantifies lithium transfer kinetics among states. State S5 shows the highest migration propensity, consistent with its coordination environment resembling bulk LPSCl, and exhibits rapid escape rates to S0 ($1.9 \times 10^{-3}$ ps$^{-1}$) and S1 ($1.6 \times 10^{-3}$ ps$^{-1}$). Notably, S4, characterized by high lithium density and weak single-anion (S/P) coordination, (**Fig. 3d**, **Supplementary Fig. 1**) enables reversible transitions to both the solid-electrolyte-like region (S5) and lithium-anode-like region (S2). As shown in **Fig. 3c**, the mobility score (normalized escape rate) identifies S2 as the kinetic bottleneck state which has highest lithium density, with the overall mobility ranking: S5 > S0 > S4 > S1 > S3 > S2. Correlation analysis between graph-derived local features and escape rates further indicates that larger average Li-neighbor distances and higher local Li density hinder fast transport, whereas the presence of anion-rich coordination (S/Cl) promotes lithium escape to other states. (**Fig. 3d**)

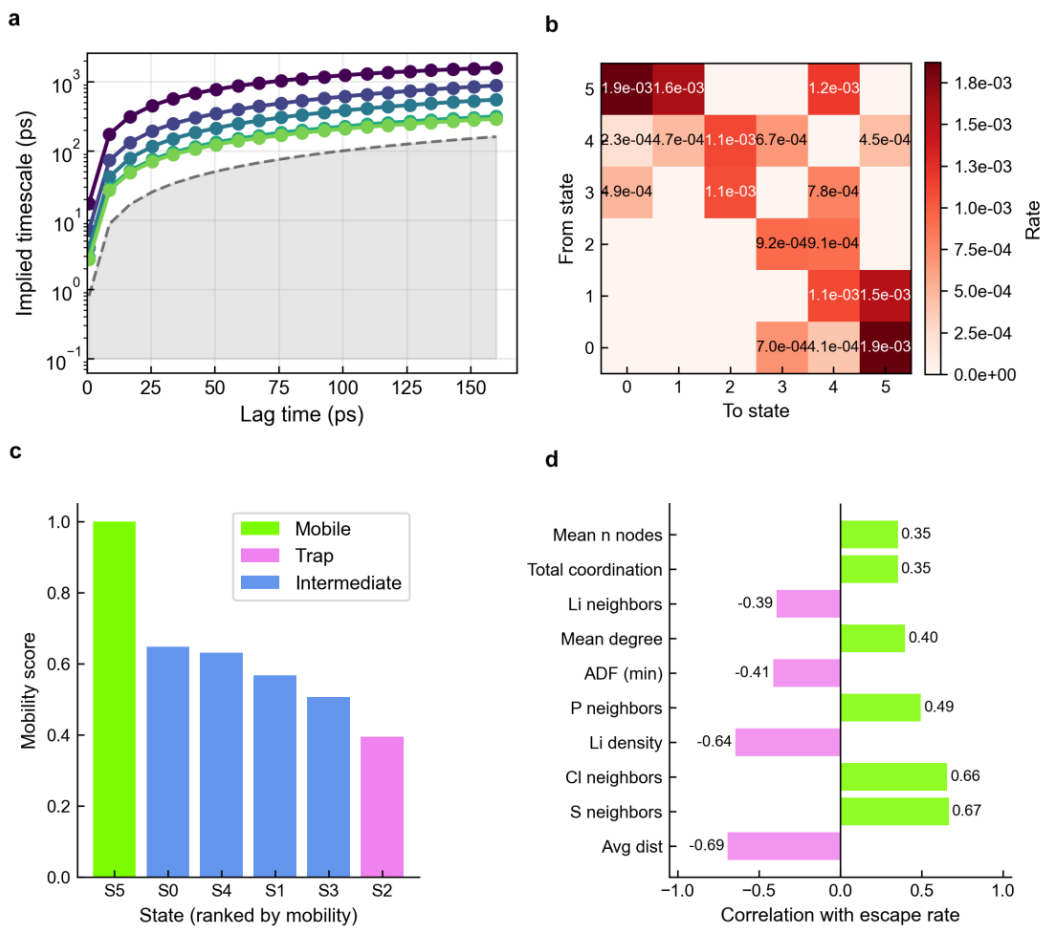

**Fig. 3 Lithium mobility analysis in LPSCl/Li. a,** Implied timescale plot derived from the Koopman process. The dashed grey line denotes lag time equals relaxation timescale. **b,** Escape rate distribution for S0–S5 lithium states. **c,** Mobility ranking of S0–S5 states in LPSCl/Li. **d,** Correlation between escape rates and node features of lithium graphs.

The escape rate matrix $R$ captures global transitions among local SEI environments, but it does not fully resolve the detailed mechanisms of lithium migration. To uncover pathway-specific transport, we quantified reactive fluxes using TPT[44,45]. As shown in **Fig. 4a**, the equilibrium flux $f_{ij}$, which represents balanced total flow independent of source-target directionality, is highest among the lithium-rich states S2, S3, and S4, consistent with active lithiation/delithiation processes in the SEI region. We then filtered $f_{ij}$ to extract transitions that originate from a specified source $A$ and terminate in a target set $B$, yielding the reactive flux $f_{ij}^+$ (**Fig. 4b-4c,** **equation 5**). Based on these reactive fluxes, we constructed the state-transition network (**Fig. 4c**). In pairwise analysis, the dominant reactive channels are S2 → S4 (associated with Li-S/P bond formation) and S2 → S3 (delithiation), confirming strong interfacial coupling between LPSCl and the lithium metal region.

Conduction routes and their associated reactive fluxes between any two states were further evaluated using a modified Dijkstra algorithm that identifies minimum edge-weight pathways (**Methods**). As shown in **Fig. 4d**, transitions from the least mobile state (S2) to the most mobile state (S5) proceed through two principal routes: (i) S2 → S4 → S5 (dominant), and (ii) S2 → S3 → S0 → S5 (secondary). These pathways suggest a common migration pattern: Li first enters a lower Li-density environment (optionally S3), is transiently coordinated by anion-rich environments (S0 or S4), and then migrates into the solid-electrolyte-like state (S5). In bottleneck routes such as S2 → S3 → S4 → S1 → S5, the S4 → S1 transition is rate-limiting. This barrier is associated with stronger multi-anion coordination of Li in high-Li-density state S1, which suppresses onward transport. These results suggest that SEI design should minimize formation of such trapping states through composition and local structure control[6].

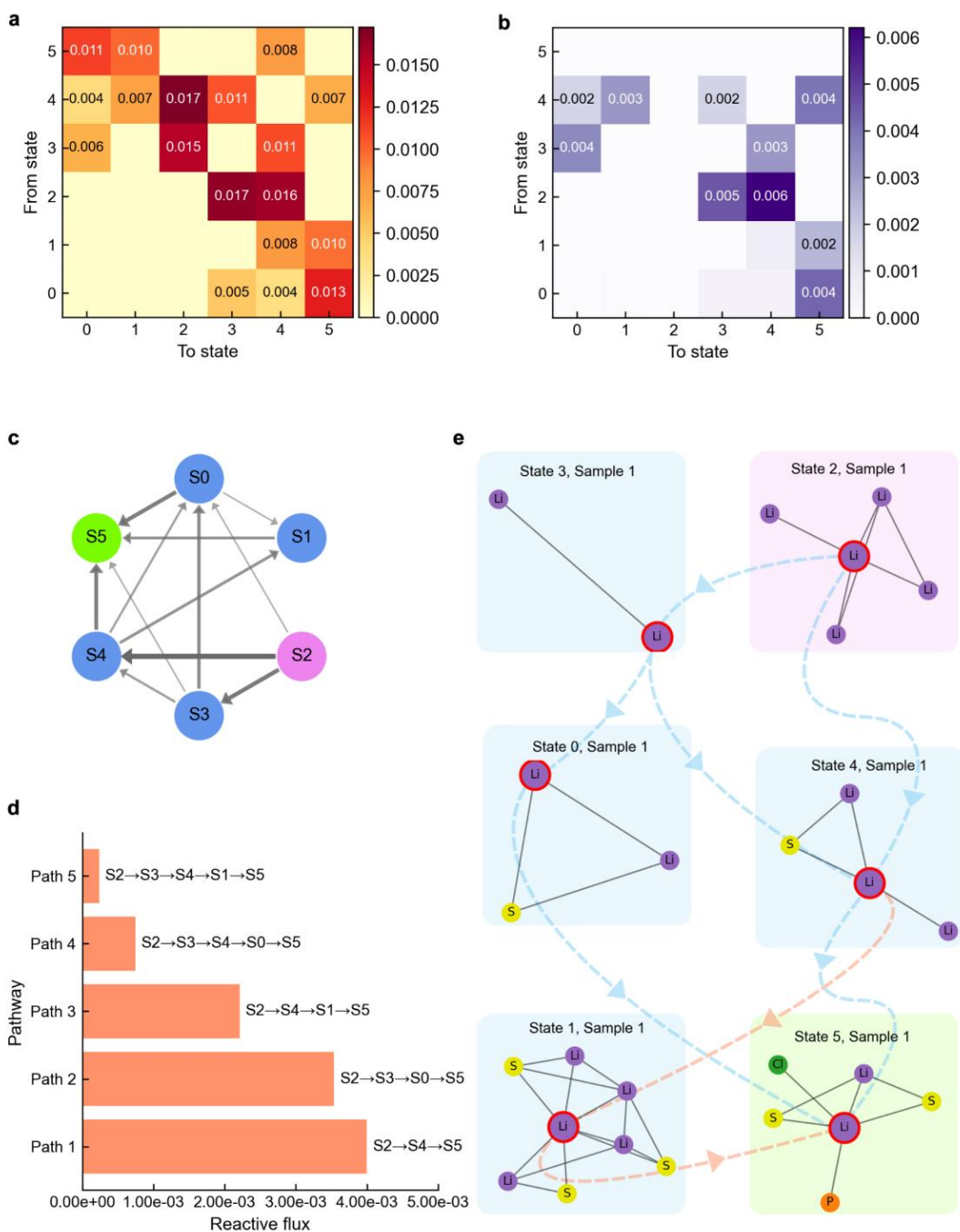

**Fig. 4 Transition pathways of lithium ions in LPSCl/Li. a-b**, Equilibrium and reactive flux between different states in the LPSCl/Li SEI. **c**, Transition network of lithium states (S0-S5). The initial and final states are colored pink and green, respectively. The width of lines represents the relative strength of the reactive flux. **d**, Top five lithium conduction pathways from S2 to S5 in LPSCl/Li SEI. **e**, Illustration of different pathways in the LPSCl/Li

SEI. Dashed blue and orange lines represent the unrestricted and restricted paths, respectively. The initial, intermediate, and final states are denoted with a pink, blue, and green background, respectively.

**Case studies in other SEI systems**

Beyond the LPSCl/Li interphase, we generalized GET-SEI to two additional interphase systems, LGPS/Li and LLZO/Li, to evaluate cross-system transferability. These systems exhibit fundamentally different interphase structures due to distinct solid-electrolyte/lithium-metal interaction. In LGPS/Li, relatively soft interfacial contact promotes formation of an intermediate SEI region spanning nanometer length scales[48,49]. (**Fig. 5a**) In contrast, LLZO/Li shows weak interfacial reactivity with lithium metal and forms a comparatively direct interface without a clearly resolved intermediate SEI layer[50,51]. (**Fig. 5b**) For both systems, dominant local environments were classified into six states (S0-S5), consistent with the LPSCl/Li setup for direct comparison (**Fig. 5c–d**, **Supplementary Fig. 2–3**), with measured Silhouette scores > 0.29 (**Supplementary Table 2**).

Reconstruction of interphase dynamics using Koopman analysis reveals markedly different lithium-transport behavior in LGPS/Li and LLZO/Li. In LGPS/Li, larger average Li-neighbor distances and higher local Li density in the SEI region correlate with lower escape rates. (**Supplementary Fig.4**) In contrast, for LLZO/Li, both increased Li density and larger average distance correlate with higher escape rates. (**Supplementary Fig.5**) This opposite trend reflects fundamentally different local-environment landscapes (**Fig. 5c-5d**): In LGPS/Li, S5 (the most mobile state; **Supplementary Fig.6**) is characterized by low Li density with anion coordination (e.g., P/S), which facilitates transitions between Li-rich anode regions and anion-rich LGPS environments. In LLZO/Li, mobile states are instead Li-rich with low oxygen content (S5) or nearly pure Li environments (S3/S1), consistent with stronger oxygen-associated trapping effects. Notably, S2 (with the lowest mobility score; **Supplementary Fig.7**) in LLZO/Li is oxygen-rich and acts as a kinetically inert barrier that suppresses lithium conduction. Koopman mode analysis further resolves five dynamical processes in LGPS/Li but only two in LLZO/Li (**Fig. 5e-5f**), reinforcing the more limited state-to-state transitions in LLZO/Li.

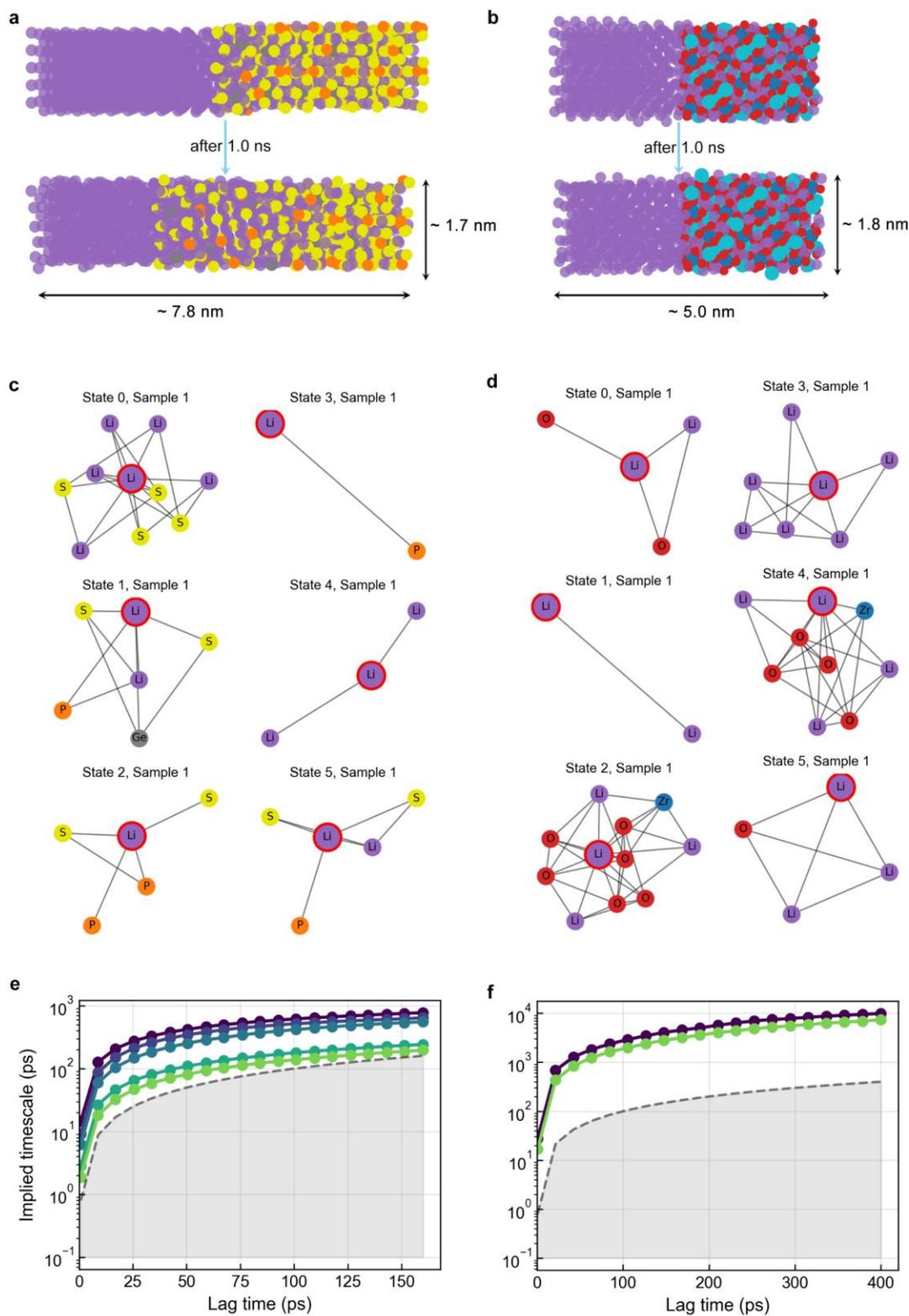

**Fig. 5 Local environments and dynamics in LGPS/Li and LLZO/Li. a-b,** Interface models of LGPS/Li and LLZO/Li employed in molecular dynamics simulations, respectively. **b-c**, Representative graphs of classified six

states in LGPS/Li and LLZO/Li SEIs, respectively. The center lithium is denoted with a red circle. **e-f**, Implied timescale plot derived from the Koopman process in LGPS/Li and LLZO/Li SEIs, respectively. The dashed grey line denotes lag time equals relaxation timescale.

Reactive lithium flux analysis further confirms the distinct lithium-transport mechanisms in these systems. As a representative comparison, we consider transitions from the least mobile state to the most mobile state (LGPS/Li: S4 → S5, LLZO/Li: S2 → S5). In LGPS/Li, transport proceeds through multiple pathways, with the direct S4 → S5 channel carrying the dominant reactive lithium flux (**Fig. 6a-6b**). Pathways involving strongly anion-coordinated intermediates (e.g., S2), such as S4→S3→S2→S5, substantially suppressed to approximately one-third of the direct flux, indicating kinetic penalties in these environments. In LLZO/Li, by contrast, conduction is dominated by a single principal route from S2 to S5, corresponding to migration from oxygen-rich to lithium-rich environments. (**Fig. 6c-6d**) Alternative pathways (Path 3-5) that pass through intermediate states (e.g., S4, S0, and S1) with progressively reduced oxygen coordination are nearly blocked, supporting the conclusion that oxygen-rich environments in LLZO/Li stabilize a kinetically inert interphase region.

Taking together, these cross-system comparisons demonstrate that GET-SEI not only resolves local lithium-environments, but also links them quantitatively to kinetic transitions and dominant transport pathways. By unifying GCL-based state discovery with EDMD and TPT flux analysis, the framework reveals how chemistry-specific local environments govern mobility, bottlenecks, and pathway diversity across sulfide and oxide interphases. These results establish a mechanistically interpretable basis for comparing interphase transport in different SEIs and provide actionable descriptors for interphase optimization.

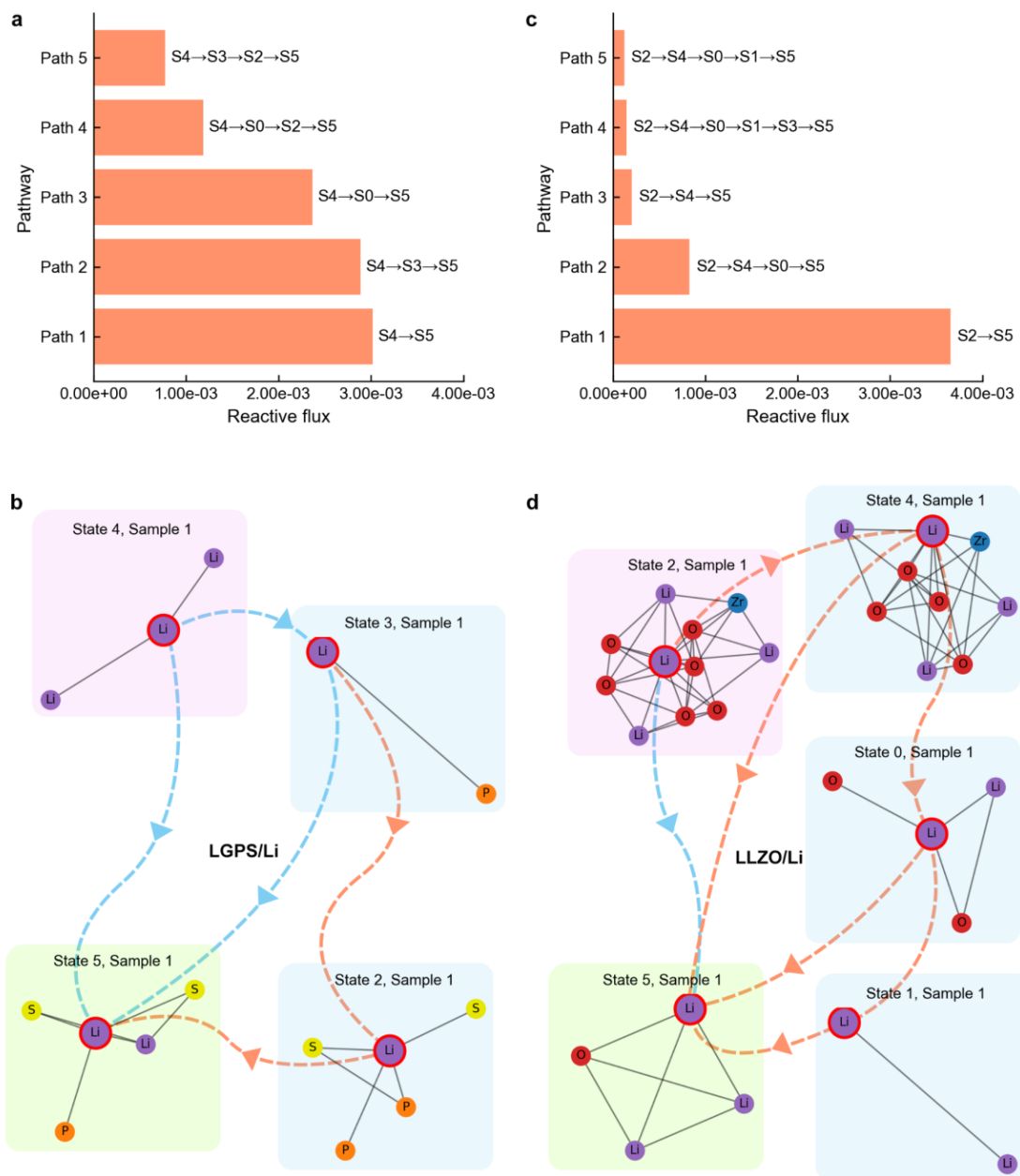

**Fig. 6 Transition pathways of lithium ions in different SEIs. a-b**, Reactive fluxes and illustration of top five dominant lithium transport pathways (S4→S5) in LGPS/Li. **c-d**, Reactive fluxes and illustration of top five dominant lithium transport pathways (S2→S5) in LLZO/Li SEI. Dashed blue and orange lines represent the unrestricted and restricted paths, respectively. The center lithium is denoted with a red circle. The initial, intermediate, and final states are denoted with a pink, blue, and green background, respectively.

**Discussion and conclusion**

Different SSEs interact with lithium metal anodes in distinct ways, leading to complex interfacial chemistries and structures. Relative to crystalline inorganic SSE phases, amorphous SEIs exhibit richer local-environment diversity and therefore more complex lithium dynamics. Quantitative characterization of lithium transport in these SEI regions is essential for rational optimization of all-solid-state battery performance. In this work, we developed GET-SEI, a general framework that uses self-supervised GCL to resolve diverse Li-centered local environments, then combines EDMD-based Koopman analysis and TPT to reconstruct nonlinear dynamics and quantify transport pathways in an interpretable way.

Applied to sulfide-based (e.g., LPSCl/Li and LGPS/Li) and oxide-based (e.g., LLZO/Li) interphases, GET-SEI quantitatively maps lithium mobility across distinct local state. Sulfide-based SEIs, especially LGPS/Li, exhibit multiple conductive pathways with higher pathway diversity, whereas LLZO/Li is dominated by oxygen-rich coordination environments that create kinetically inert barriers and suppress transitions. These results suggest a practical design principle for fast interfacial transport: increasing the connectivity and population of high-mobility states while suppressing strongly coordinated trapping states. More broadly, GET-SEI provides a transferable, mechanism-informed metric framework for evaluating and designing SEIs toward fast, stable lithium transport in all-solid-state batteries.

**Methods**

**Neural-network potential preparation**

All the neural-network potentials (NNPs) were developed using MACE foundation model[52]. Initial data was prepared through *ab initio* molecular dynamics (AIMD) simulations using CP2K[53]. The Perdew–Burke–Ernzerhof (PBE) generalized gradient approximation (GGA)[54] with the double-zeta valence polarized basis set and Goedecker–Teter–Hutter pseudopotentials were adopted[55]. The auxiliary plane-wave basis was truncated at a density cutoff of 500 Ry, and van der Waals interactions were treated with the Grimme D3 correction. A Nosé-Hoover thermostat and Martyna-Tobias-Klein barostat were used, each with a coupling constant of 1.0 ps. The time step was 2.0 fs.

For the LPSCl/Li system, six initial structures containing $Li_2S$, $LiCl$, and $Li_3P$ units were generated with Packmol[56] based on experimental findings[15], together with LPSCl[57] and an explicit Li/LPSCl interfacial model. Each system was simulated by 10.0 ps of AIMD at 300-1500 K and 1 atm in the NPT ensemble. Two frames from each NPT trajectory were then selected for follow-up NVT simulation with multithermal sampling using on-the-fly probability method[58]. The final dataset comprised 28648 training configurations and 7162 test configurations. The final NNP was obtained by further fine-tuning the mace-mpa-0-medium model[52]. Four trained models were evaluated, with accuracy metrics reported in **Supplementary Table 1**. For other SEI systems (LGPS/Li and LLZO/Li), the mace-mpa-0-medium model was directly employed for simulations.

**Molecular dynamics simulation**

All NNMD trajectories were generated through the ASE package[59]. A timestep of 2.0 fs was used for all simulations. All NNMDs employed a 2.0 fs timestep. For each system, a 1.0 ns NNMD simulation under NVT ensemble at 300 K was performed with a Langevin thermostat (coupling constant: 1.0 ps). Cell parameters and atom counts are summarized below: (1) LPSCl/Li: 27.04 × 27.92 × 95.22 Å, including 2960 atoms. (1) LGPS/Li: 17.38 × 17.20 × 96.95 Å, including 1200 atoms. (3) LLZO/Li: 18.47 × 18.48 × 71.79 Å, including 1167 atoms.

**Graph construction**

Individual lithium graphs were created through Pytorch geometric[60]. The graph edges were constructed through cutoff-based algorithm using a cutoff of 3.0 Å. Four kinds of node features were selected: (1) Atomic number, (2) Average distance between the center lithium atom and neighbor nodes, (3) Number of each element in the local coordination shell (a larger shell radius of 4.0 Å is used), (4) Summary of statistics of angular distribution functions (ADFs) were used to describe the positional distribution of the atoms, including values of mean, standard deviation, minimum, maximum, and median of the ADFs. The ADFs were calculated by computing angles for all unique pairs of neighbors within the cutoff distance of 4.0 Å.

**Graph contrastive learning architecture**

The input of the GCL is lithium graphs generated from NNMD trajectories. 20% of the NNMD frames were used as training data for the local environment classification. For the augmented views of Li graphs, the 0.1 edges were randomly dropped out, and 0.2 node features were randomly masked. The temperature σ was selected as 0.5 for a good balance between the sharp and smooth embeddings. After data augmentation, they were fed into the graph attention network (GAT) encoder, which includes 2 layers with multi-head attention (4 heads), learning the neighbors surrounding the Lithium atom. Then the output was projected to a 32-dimensional embedding space. Then we clustered the embedding space through Gaussian mixture model through sci-kit learn package[61]. Finally, we assigned states to each Li at each frame in our simulated trajectories.

**Construction of Koopman operator through EDMD**

EDMD approximates the Koopman operator using a dictionary of basis functions $\{\psi_1, \psi_2, \cdots, \psi_D\}$. Common dictionaries include monomials, radial basis functions and through neural networks like VAMPnets[27,29]. Here, since we already have discrete states for lithium atoms after graph contrastive learning, we use the indicator functions through one-hot encoding which is efficient and simple:

$$\psi_i(x) = \begin{cases} 1, & x = i \\ 0, & x \neq i \end{cases} \tag{6}$$

Then we collected the data pairs $(x_n, x_{n+1})$ from the NNMD trajectories. These data pairs were transformed into an observable space $(g(x_n), g(x_{n+1}))$, with each expressed by the dictionary functions $\psi$:

$$g(x_n) = [\psi_1(x_n), \psi_2(x_n), \cdots, \psi_D(x_n)] \tag{7}$$

Through solving the least squares problem in the observable space, we could have the finite-dimensional matrix approximation $K$ of $\mathcal{K}$:

$$\mathcal{K} \approx K = (G_n^T G_n)^{-1} G_n^T G_{n+1} \tag{8}$$

Where $\boldsymbol{G}_n$ and $\boldsymbol{G}_{n+1}$ are matrices form of $g(x_n)$ and $g(x_{n+1})$, whose rows are the observables at time $n$ and $n+1$ respectively.

**Algorithm in finding dominant pathways**

The dominant pathways between source and target states were identified using TPT combined with a greedy flux decomposition. After having the reactive fluxes $f_{ij}^+$ (**equation 5**), we decomposed it into individual pathways using an iterative algorithm: at each iteration, a modified Dijkstra search finds the path from source to target with the maximum bottleneck flux, that path's bottleneck flux is subtracted from all its edges, and the procedure repeats on the residual flux network until the desired number of pathways is extracted. Finally, we have the results of an ordered set of dominant pathways ranked by their flux contributions, providing a mechanistic decomposition of how reactive transitions flow through the intermediate states of the SEI system.

**Acknowledgments**

This work was performed in part at the High-Performance Computing Cluster (HPCC) which is supported by the Information and Communication Technology Office (ICTO) and Super Intelligent Computing Center (SICC) of the University of Macau.

58  Invernizzi, M. & Parrinello, M. Rethinking Metadynamics: From Bias Potentials to Probability Distributions. *J. Phys. Chem. Lett.* **11**, 2731–2736 (2020).

59  Hjorth Larsen, A. *et al.* The atomic simulation environment-a Python library for working with atoms. *J. Phys. Condens. Matter.* **29**, 273002 (2017). https://doi.org/10.1088/1361-648X/aa680e

60  Fey, M. & Lenssen, J. E. Fast graph representation learning with PyTorch Geometric. *arXiv preprint arXiv:1903.02428* (2019).

61  Pedregosa, F. *et al.* Scikit-learn: Machine learning in Python. *J. Mach. Learn. Res.* **12**, 2825–2830 (2011).
25

*Supplementary data*

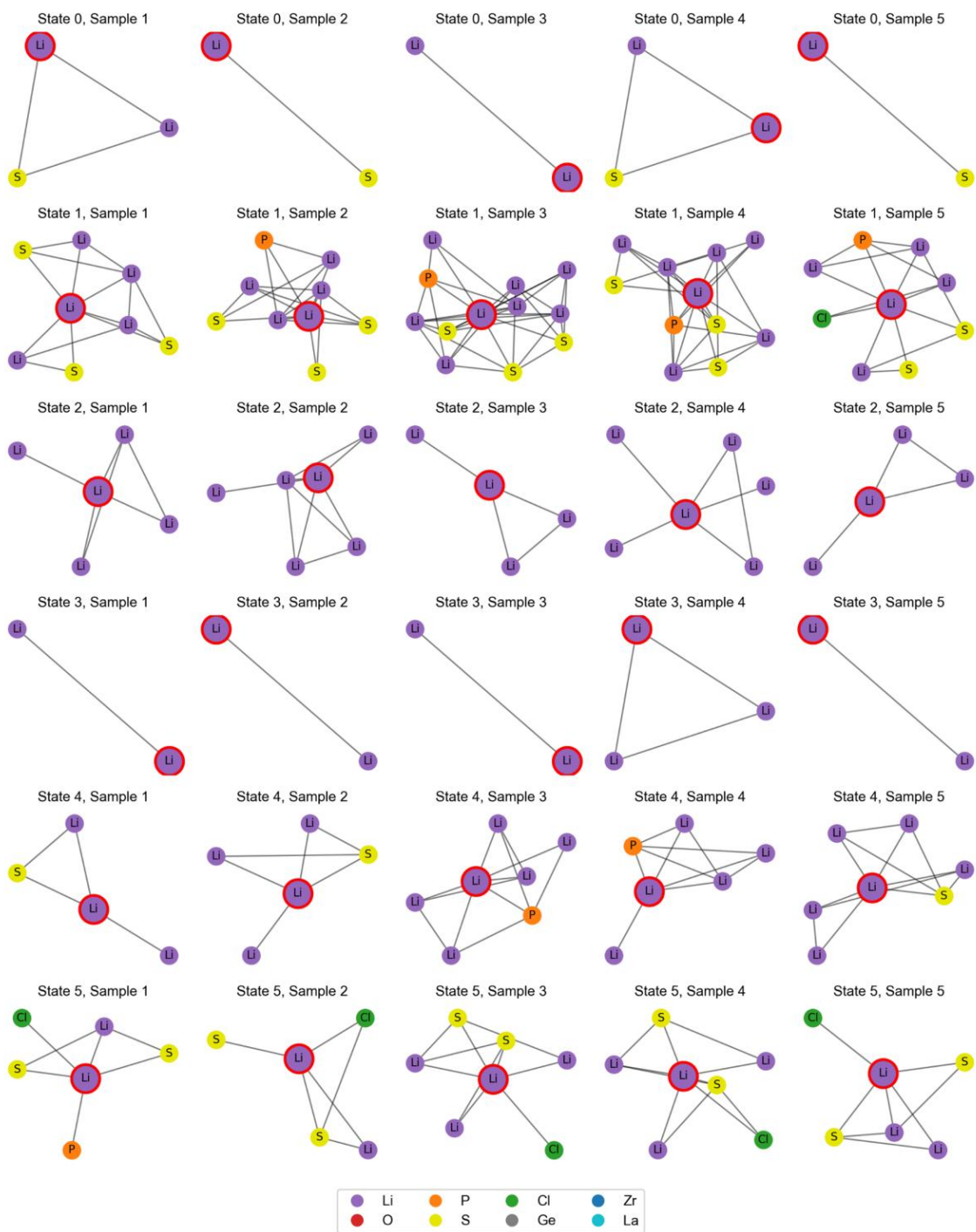

**Supplementary Fig. 1** Five representative graphs of each classified state in LPSCl/Li.

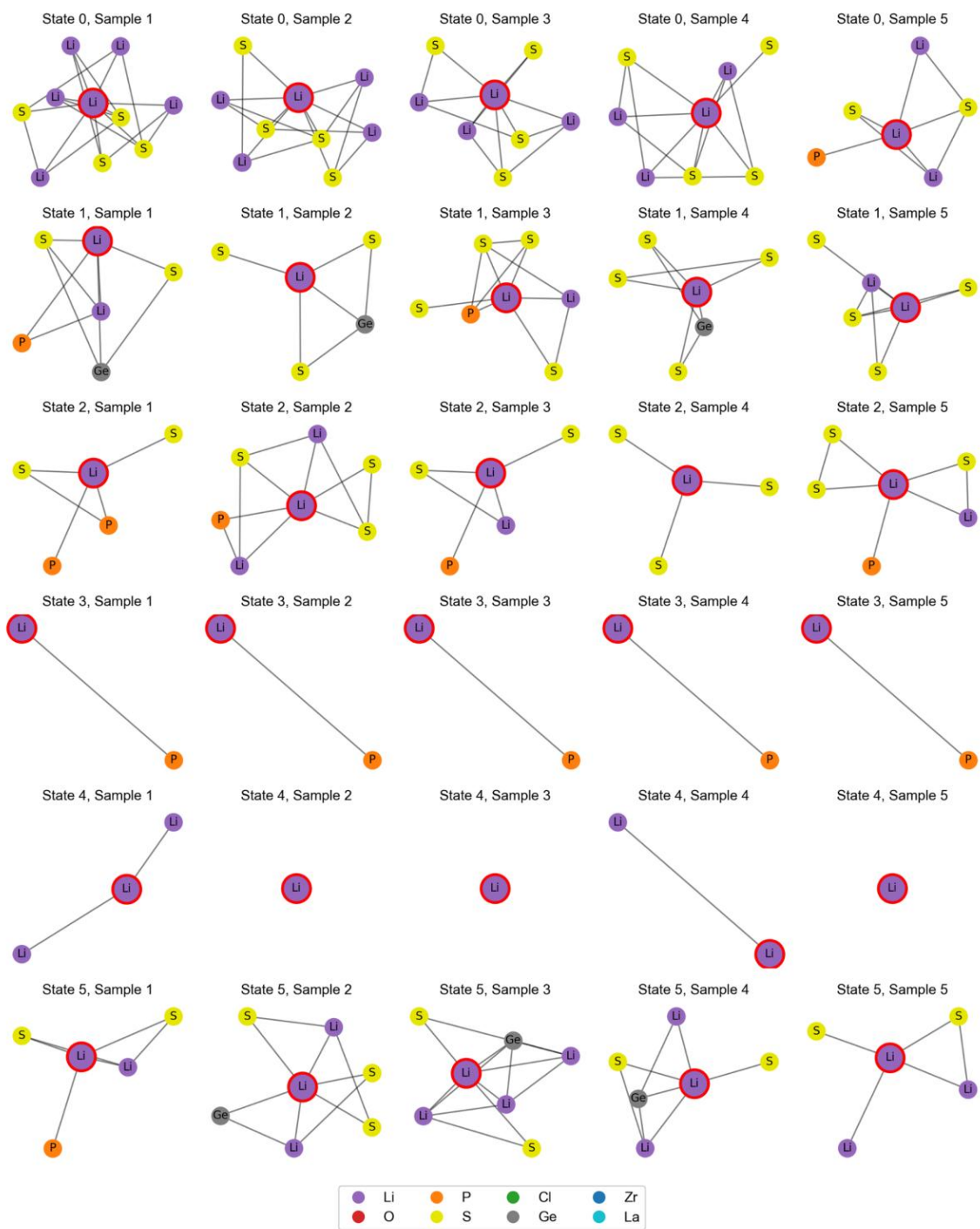

**Supplementary Fig. 2** Five representative graphs of each classified state in LGPS/Li.

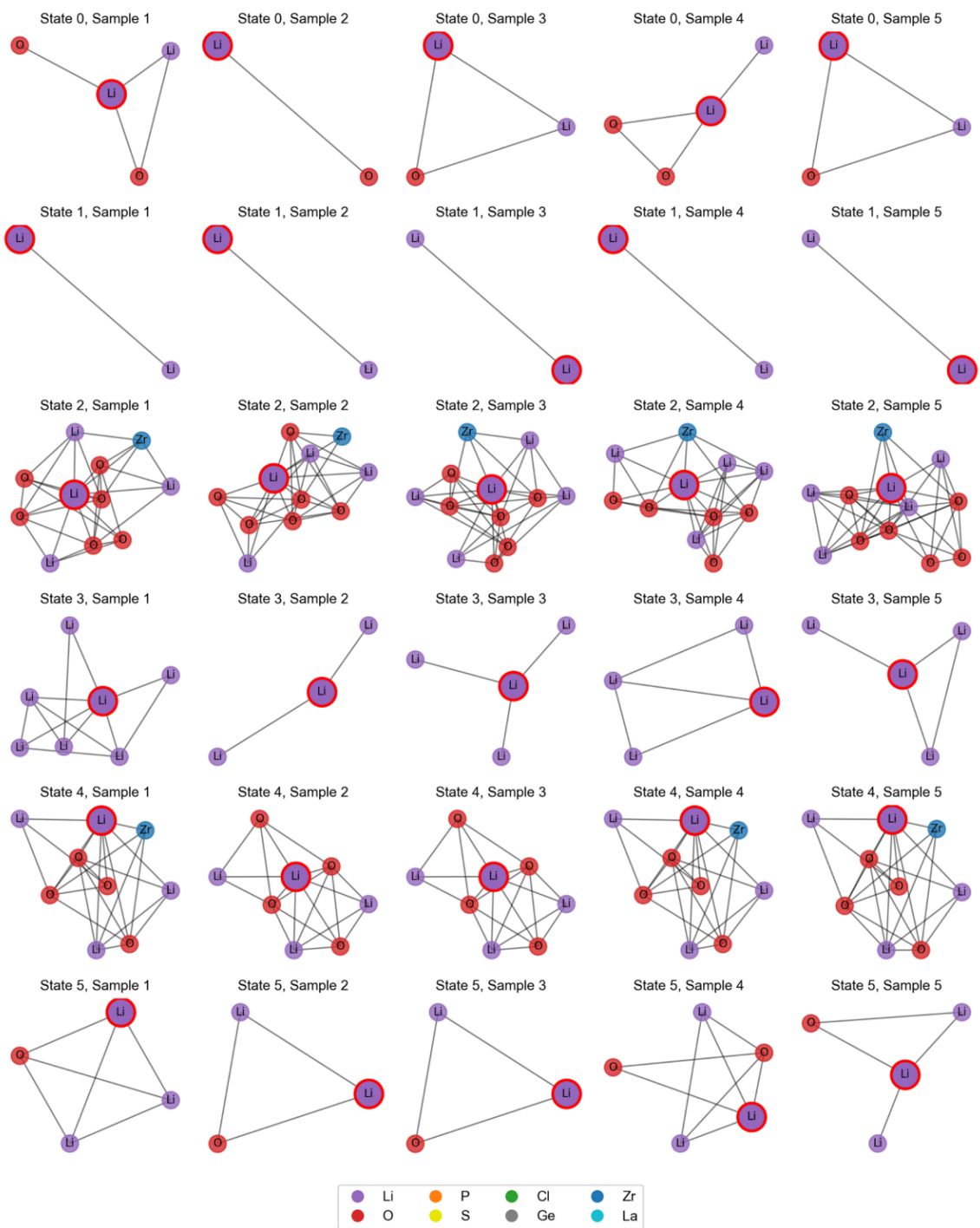

**Supplementary Fig. 3** Five representative graphs of each classified state in LLZO/Li.

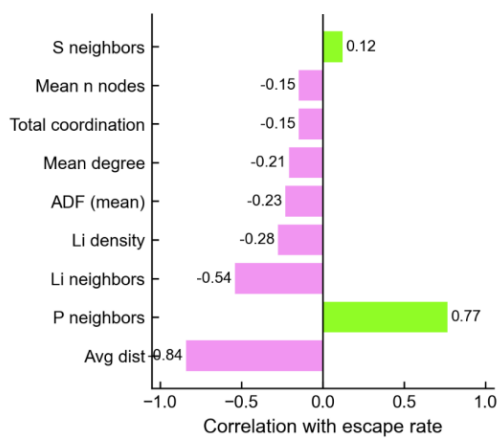

**Supplementary Fig. 4** Correlation of escape rate with node features of Li graph in LGPS/Li.

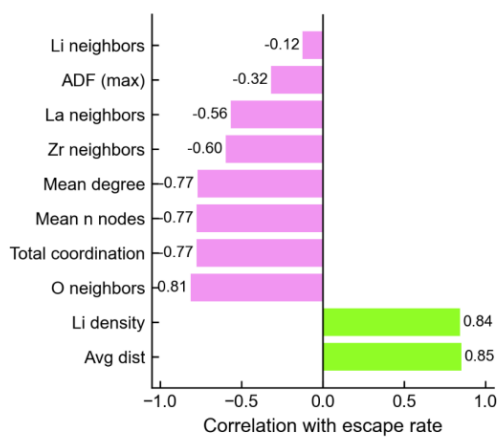

**Supplementary Fig. 5** Correlation of the escape rate with node features of Li graph in LLZO/Li.

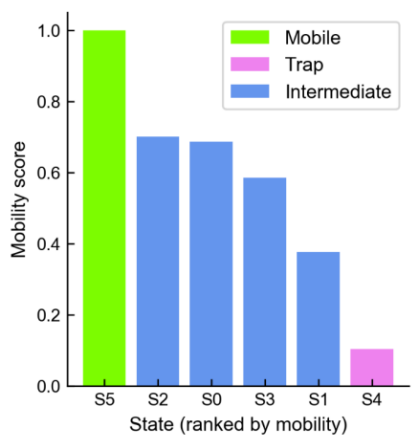

**Supplementary Fig. 6** Mobility ranking of S0–S5 states in LGPS/Li.

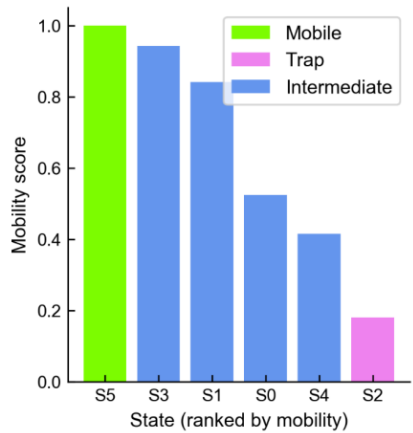

**Supplementary Fig. 7** Mobility ranking of S0–S5 states in LLZO/Li.

**Supplementary Table 1**. Root mean square errors (RMSEs) of trained neural-network potentials (LPSCl/Li).

|         | Energy (meV/atom) | Force (meV/Å) |
|---------|-------------------|---------------|
| Model-1 | 0.6               | 21.4          |
| Model-2 | 0.6               | 21.8          |
| Model-3 | 0.6               | 21.2          |
| Model-4 | 0.6               | 21.5          |

**Supplementary Table 2**. Silhouette score of classification through gaussian mixture model in different SEI systems.

|                  | LPSCl/Li | LGPS/Li | LLZO/Li |
|------------------|----------|---------|---------|
| Silhouette score | 0.2852   | 0.2974  | 0.3300  |

**Supplementary Note 1**

Silhouette score ($s$) was used to measure the cluster quality of lithium graphs. It was calculated through the scikit-learn package[1]:

$$s = \frac{(b - a)}{\max(a, b)}$$

Where $a$ is the average intra-cluster distance, and $b$ is the average nearest-cluster distance. Commonly, $s > 0.25$ indicates a reasonable classification.

**Supplementary References**